# The seismogenic area in the lithosphere considered as an "Open Physical System". Its implications on some seismological aspects. Part - I. Accelerated deformation.

Thanassoulas, P.C., B.Sc in Physics, M.Sc – Ph.D in Applied Geophysics.

Retired from the Institute for Geology and Mineral Exploration (IGME)
Geophysical Department, Athens, Greece.
e-mail: thandin@otenet.gr – website: www.earthquakeprediction.gr

**Abstract.**

The seismogenic area in the lithosphere is considered as an open physical system. The mathematical analysis of its inflow – outflow energy balance reveals the presence of specific energy flow physical models. The later comply with what has been observed by seismologists as "accelerated deformation" and "seismic quiescence". These physical models are represented by low-order cumulative seismic energy release analytical polynomial functions in time. The time derivatives of these functions, analytically calculated, provide a more sharp indication for the time of occurrence of a future large EQ. Examples from the application of this approach on real large EQs from the Greek territory are presented. Moreover, this methodology can be used for the determination of: the maximum expected magnitude of a future large EQ of a specific seismically activated seismogenic area and the compilation of seismic potential regional maps. The later will be presented in details in articles to follow soon.

## 1. Introduction.

The issue of the earthquake prediction has raised strong debates in the scientific community. One of the arguments, which are presented against it, is the "absence of a physical model" that will justify any methodology to this end. In other words, there is no physical mechanism, universally agreed, for the seismogenesis.

The different physical models, which have been proposed up to date, such as the "Rebound Model" (Reid, 1911), the "Self Organized Criticality (**SOC**) model (Bak et al. 1988, Bak 1996), the Chaotic Non-linear Systems (Anderson 1990, Kagan 1994, Main 1996), highly depend on the initial conditions, thus suggest the unpredictability of EQs. Other studies have shown that a small EQ can grow into a strong one, depending on very small variations on elastic properties, fault parameters variations as friction, stored strain energy (Otsuka 1972a, b; Vere-Jones1976, Ito et al. 1990).

Moreover, the fault friction plays the most important role in seismogenesis, in the frame of the tectonic regime of the lithosphere, while fracturing is a secondary one (Scholz, 1998). The term "stick-slip frictional instability", introduced by Brace and Byerlee (1966), suggests that strain energy is accumulated during the "stick" period, while an EQ occurs at the "slip" period. Actually, the phenomenon of seismogenesis is a mixture of frictional slip failure and shear fracture (Ohnaka, 2003).

The seismogenesis mechanisms, postulated to date, refer to the physical-mechanical procedures that take place in the focal area, before and during an EQ occurs. In a most indirect way they refer to EQ precursors and therefore, to the capability of prognostic parameters determination. An example of such a calculation is the "power law - time to failure model" that correlates the magnitude (**M**) of a future strong EQ to the remaining time ($t_r$) towards its occurrence (Bufe and Varnes 1993, Bowman et al. 1998).

The absence of a valid and robust relation between the different seismogenic mechanisms and the seismic precursors, required for the determination of the prognostic parameters (location, magnitude, time), led the seismologists to apply statistical methods for the issue of the earthquake prediction. Methodologies, such as the algorithms **M8** (Keilis-Borok et al. 1990, Healey et al. 1992, Romashkova et al. 2002) **and CN** (Keilis-Borok et al. 1990a), have been applied with some success, mainly for long-term and medium-term prediction for rather large areas.

At this point it must be stressed out that, what is really needed for a successful earthquake prediction is <u>**a physical model upon which the appropriate calculations will be based in order to analyze the precursory available data,**</u> so that earthquake prediction can be implemented. The difference between these physical models and the physical, seismogenic mechanisms is essential. The seismogenic mechanisms refer to what actually happens in the seismogenic region and therefore, they are a physical "close-up" view in the seismogenic area, while the physical models, used, for the calculation of the prognostic parameters, are a more generalized, physical approach of the seismogenic region, and they are independent from the actual seismogenic mechanisms which take place in it.

Moreover, it is anticipated that a single physical mechanism cannot provide answers for all earthquake prognostic parameters. This is evident from the large number of publications, which are related to the topic of earthquake prediction. The majority of them refer only to time, or regional area in relation to magnitude. None of them deals, simultaneously, with all the prognostic parameters.

In this presentation, the seismogenic area will be considered from the physics point of view. Particularly, it will be treated as an "open physical system". In other words, the seismogenic area can absorb stress energy from its surrounding region and simultaneously is capable to release energy through its seismic activity into its surrounding. This kind of model is called "Lithospheric Seismic Energy Flow Model (**LSEFM**)".

Actually, this derives from the direct application of the energy conservation law of physics. That is "the input energy in the model must equal to the released plus the stored one in it". This physical concept explains, in energy transfer terms through the lithosphere, the "accelerated deformation" and "seismic quiescence" methodology, used by the seismologists and facilitates the calculation of the maximum expected magnitude of an imminent, strong EQ, while an analytical determination can be made for the seismic potential of a seismogenic large area.

In this work (**Part – I**), the "Accelerated Deformation" will firstly be considered.

## 2. Theoretical analysis.

It is generally accepted that stress energy built-up, in a focal area, is a very slow process which, closely, follows the motion of the lithospheric plates. It takes a long period of time (probably a large number of years) to reach the point when an earthquake will occur, because of rock collapsing.

Under normal conditions, the stored energy is discharged through the background, small magnitude, seismicity of the seismogenic area. The later is demonstrated in figure **(2. 1)**.

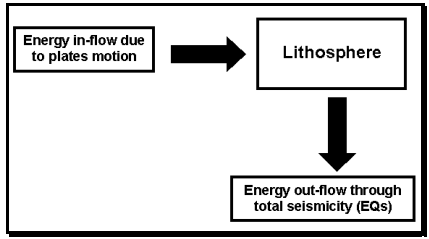

Fig. 2. 1. The postulated model of seismic energy flow, through a seismogenic area of the lithosphere.

When a strong earthquake is in preparation, before its occurrence, the normal seismicity either decreases for a certain period of time and therefore, the seismic "quiescence", which is detected, is used as a precursory indicator or increases and therefore, the "accelerated deformation" is observed.

The mathematical analysis of the postulated model is as follows:

Let us denote as **(Ein)** the inflow energy over a short period of time (**dt**) in a seismogenic area of the lithosphere and as **(Eout)** the energy outflow of the same area, due to its seismicity for the same period of time. **(Ein)** and **(Eout)** are time functions, **Ein = Ein(t)** and **Eout = Eout(t)**.

The seismogenic area is charged with energy **Est = Est(t)** as follows:

$$Est = Ein - Eout \qquad (2.1)$$

For any very short period of time **(dt)**, equation **(2.1)** takes the following form:

$$Est(dt) = d(Est(t)) \qquad (2.2)$$

For successive values of **dt**, a discrete valued function **Y(t)** is defined as follows:

$$Y(t) = Est(t) \qquad (2.3)$$

Equation **(2.3)** can take the following forms:

a. $Y(t) = C = 0$      (2.4)

In this case, the stored energy in the seismogenic area, equals to zero, therefore, it is at a state of stable zero-charge conditions.

b. $Y(t) = C < 0$      (2.5)

In this case, the stored energy decreases, therefore, the seismogenic area continuously discharges, towards a very stable state of uncharged conditions.

c. $Y(t) = C > 0$      (2.6)

In this case the stored energy increases continuously, therefore, the seismogenic area is charged towards a state of highly unstable conditions, leading to the occurrence of an earthquake.

The term (**C**) of equation **(2.6)** can generally be either time dependent or time independent.

- <u>**C** is time independent:</u>

Equations **(2.2)** and **(2.6)** can be combined in equation **(2.7)**.

$$d(Est(t)) = C > 0 \qquad (2.7)$$

The cumulative energy (**Ecum**), stored in the seismogenic area, can be calculated as a function of time (**Ecum(t)**) by integrating both sides of equation **(2.7)**, in respect to time **(t)**,

$$Ecum(t) = \int d(Est(t))dt = \int Cdt \qquad (2.7a)$$

Since term **C** is constant and time independent, the calculated function **Ecum(t)** has the form of:

$$Ecum(t) = C*t + b \qquad (2.8)$$

Where, **b** denotes the integration constant.

The linear equation **(2.8)** was firstly introduced by Thanassoulas et al. (2001), along with the postulated, lithospheric seismic energy flow model, for the calculation of the maximum expected magnitude of an imminent, strong EQ.

- <u>**C** is time depended:</u>

In this case, equation **(2.7)** can be represented by an $n^{th}$-order polynomial:

$$d(Est(t)) = a_n t^n + a_{n-1} t^{n-1} + \ldots\ldots + a_0 \qquad (2.9)$$

The cumulative energy **Ecum(t),** stored, in the seismogenic area, can be calculated as a function of time by integrating in time, both sides of equation **(2.9).**

$$\text{Ecum}(t) = \int d(\text{Est}(t))dt = \int ( a_n t^n + a_{n-1} t^{n-1} + \ldots + a_0 )dt \qquad (2.9a)$$

As an n+1 order polynomial,

$$\text{Ecum}(t) = k_{n+1} t^{n+1} + k_n t^n + \ldots k_0 \qquad (2.9b)$$

Where, $k_i$ values represent the polynomial constants.

Summarizing the forms the equation **Ecum(t)** (that represents the energy flow) takes, the following cases are possible:

**a. linear polynomial  -  constant energy flow**

**b. higher order polynomial  -  real acceleration**

**c. <u>accelerated</u> for a period of time long before the main, seismic event, which is followed by a <u>constant energy flow</u>, just before the occurrence of the strong EQ.**

Cases **(b)** and **(c)** are represented, very often, by the well-known "time to failure" function.
Although, mathematically, it is generally possible to transform any polynomial to any arbitrary function, i.e. time to failure function, by calculating with LSQ techniques the appropriate transformation parameters of the later, still remains the issue of arbitrariness, as far as it concerns the validity of physics behind this transformation.
Moreover, the time to failure function depends on two variables. The first one is the magnitude of the imminent EQ and the second one is the time to failure left. In order to overcome the problem of solving a two parametric equation (of infinite number of solutions), the parameter **C** (Bowman et al. 1998) was introduced, that is the ratio of power low fit error over the linear fit error, as far as it concerns the cumulative seismic energy release. Still, the notion of this ratio is set completely, arbitrarily.
Therefore, it is suggested that the magnitude and time to failure of a strong, imminent EQ, calculated, by these methodologies, are not supported by any validated, physical mechanism and should rather be rejected.

## 3. Application of the model on real EQ cases.

The application of the lithospheric, seismic energy flow model requires two parameters to be known, in advance. The first one is an initial estimation of the area extent of the open physical system itself (seismogenic area). The second one is its seismic history.
The choice of the first parameter is utilized either by (a) knowing the epicentral area of the future EQ by another methodology, (b) by taking into account the increased low level seismicity of an under study seismogenic area, which is some times observed and finally (c) by taking into account the deep, lithospheric fracture zones which are mapped by the conversion of the regional gravity map of the seismic region in to a gradient one (Thanassoulas, 1998).
This procedure is explained by the figures to follow.
As long as the seismogenic area has been activated, a small magnitude seismic activity, associated with the main rupture of the rock formation, generally increases in the different order fracturing branches (n$^{th}$ order ridel) and therefore, this seismic activity is observed along and very close to the trace of the main seismic fault to be activated. The later is presented in the following figure **(3.1)**.

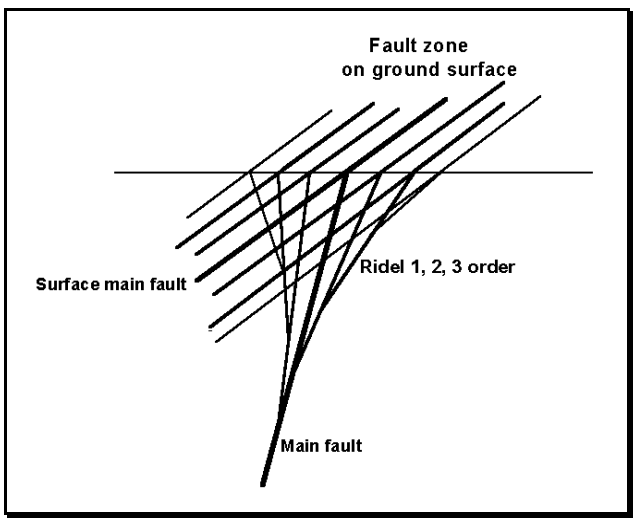

Fig. 3.1. The main, seismic fault branches into smaller ones, as it approaches the ground surface (Mattauer 1973, Vialon et al. 1976). The group of fractures that advances up to the ground surface forms the observed "fracture zone".

Since small-scale seismicity has started to evolve, the corresponding epicenters of the foci will be located close to the vicinity of the main fault, which will be activated in the future. A sketch drawing indicating this main earthquake precursory activity is presented in the following figure **(3.2)**.

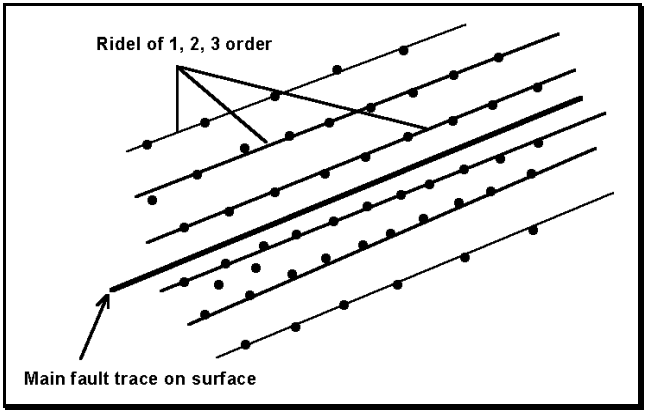

Fig. 3.2. Sketch drawing of the theoretically small magnitude seismicity, expected, to be observed on ground surface and in the vicinity of a main, seismic fault, prior to a strong EQ. The solid dots represent the epicenters of the precursory, seismic activity, while the thin lines indicate the $n^{th}$ order "ridel" different faults. The thick, black line represents the main fault which is going to be activated.

Consequently, the physical system which will be studied must be chosen, close, to the main fracture, in a way that it takes into account this specific distribution of the small magnitude EQs. The second parameter, the seismic history of the seismogenic area, is taken from the EQ files from the Seismological State Observatories. In the case of Greece, this is the National Observatory of Athens (NOA). The application of the methodology is shown in the following figures. It is assumed that, the epicentral area has been approximately determined by other means. As a first step, the seismogenic area is defined which will be considered as the corresponding open physical system of which the seismic energy release will be analyzed **(fig. 3.3)**.

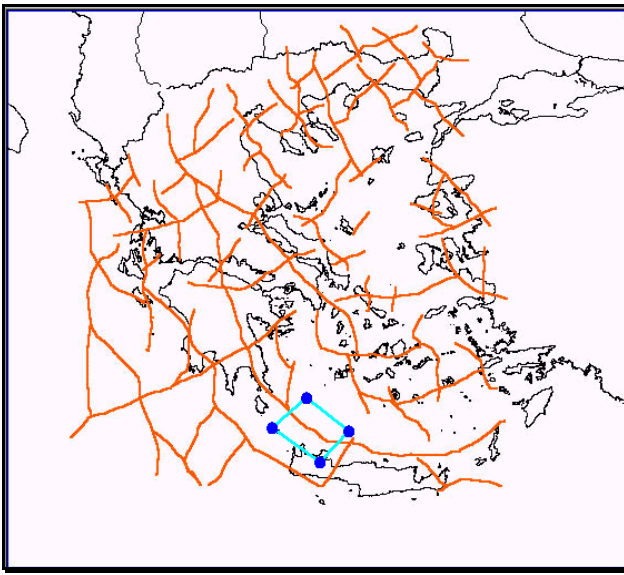

Fig. 3.3. The approximately, determined seismogenic area is indicated by the blue polygon. It has been drawn in such a way to center the associated deep, main, lithospheric fracture zone and is elongated along it.

The next step to take is to calculate backwards the cumulative energy release of this area for a long period of time. The graph which results indicates whether this area has entered the acceleration phase or not. The later is presented in the following figure **(3.4)**.

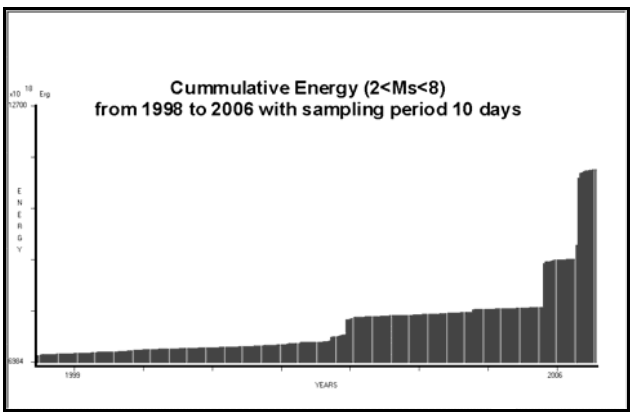

Fig. 3.4. Cumulative seismic energy release calculated, for the seismogenic area, which is indicated in the previous figure **(3.3)** for the period 1998 – 2006.

The study of this graph indicates that this sample, seismogenic area has been set into cumulative, seismic energy release acceleration mode for the last 8 years. A slightly different presentation is shown in next figure **(3.5)** by fitting a $6^{th}$ degree polynomial. This facilitates the analytical calculation of the time gradient of the cumulative seismic energy release.

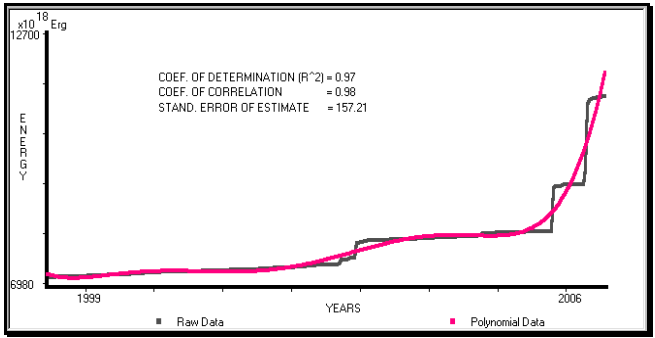

Fig. 3.5. Cumulative, seismic energy release fits with a 6<sup>th</sup> degree polynomial. The black line indicates the cumulative, seismic energy release data, while the red one indicates the fitted, polynomial values.

It is obvious that the acceleration has been initiated since the end of 2005, indicating that an earthquake will strike "soon". Actually, a strong (**M = 6.9R**) earthquake took place at a short distance towards NW on this lithospheric fracture zone at the start of 2006. Its details will be presented in the examples to follow.

Quite often, the start of the increase of the cumulative energy release is not defined very sharply, but there is a gradual change over a rather lengthy period. In such cases, instead of fitting only a polynomial, as an advanced step, the corresponding time gradient of the polynomial is calculated analytically.

The process of time gradient calculation acts as a high-pass filter on the cumulative seismic energy release data, low-order terms of the fitted polynomial are excluded by this operation and therefore the resolving capability for detecting smaller level changes is larger.

This process is shown in the following figure **(3.6)**.

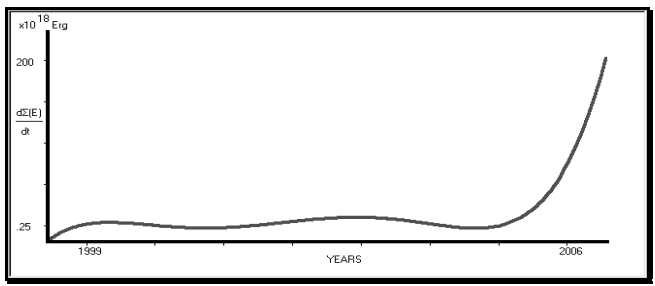

Fig. 3.6. Cumulative, seismic, energy release time-gradient, calculated analytically, from the polynomial of figure **(3.5)**.

### 3.1. Application of the theoretical model on the EQ of Zakynthos (02/12/2002, Ms = 5.8R).

The first example, based on the methodology which has been already presented, is that of Zakynthos EQ. A few months before the main, seismic event, which took place on 02/12/2002 with a magnitude of Ms = 5.8R, an increased, small-scale, seismic activity had been observed at Zakynthos area. The area of interest is indicated by a red circle in the following figure **(3.1.1)**.

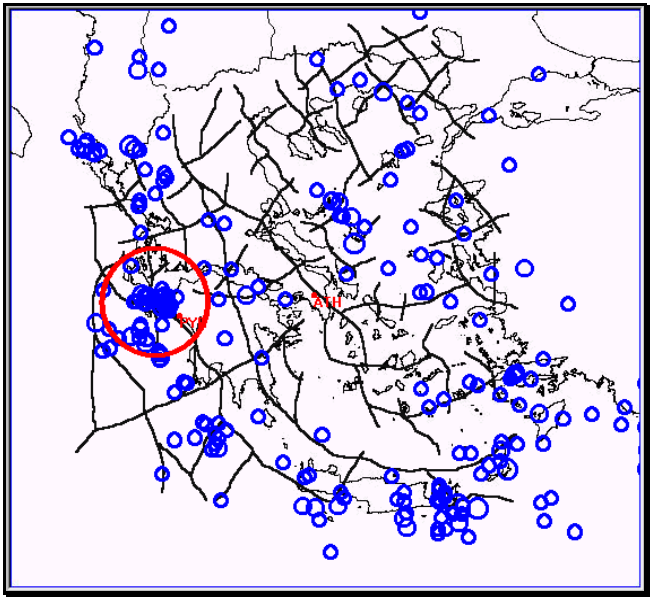

Fig. 3.1.1. The red circle indicates the regional area of Zakynthos. The blue circles indicate the small magnitude seismic activity all over Greece for a time period of a couple of months before the EQ occurred.

The seismicity which was observed to increase, dictated the application of the proposed methodology. The area, to be investigated, was selected in such a way, to include the main, lithospheric fracture zones, which are present in the region. The later is presented in the following figure **(3.1.2)**. The blue frame indicates the area of interest.

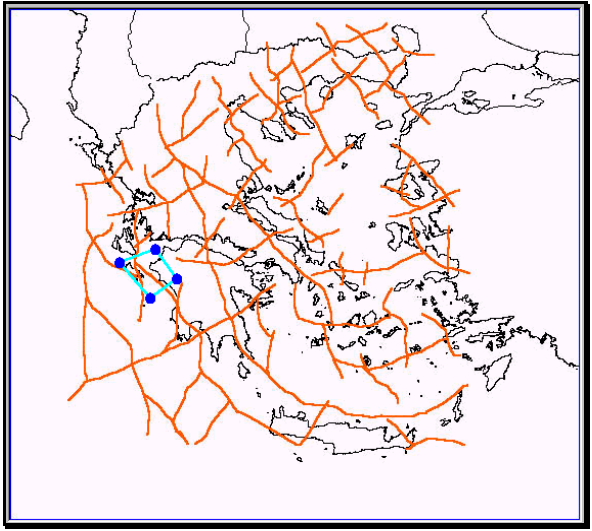

Fig. 3.1.2. Area of interest (blue frame) for which the lithospheric seismic energy release methodology will be applied. The brown lines indicate the deep lithospheric fracture zones.

Next step is to calculate backwards the cumulative seismic energy release for some years. The later was utilized by using the EQ file, which is available online in the web by the NOA, Athens, Greece. This operation indicated that the area of Zakynthos was set in seismic acceleration mode from the start of the year 2000. This is presented in the following figure **(3.1.3)**.

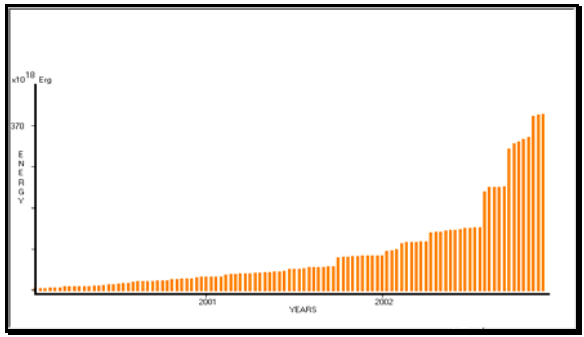

Fig. 3.1.3. Cumulative seismic energy released, from the seismogenic area of Zakynthos for the period 2000 – 2002.

The 6$^{th}$ order polynomial, fitted in the cumulative, seismic energy release data for the area of Zakynthos, is presented in the following figure **(3.1.4)**.

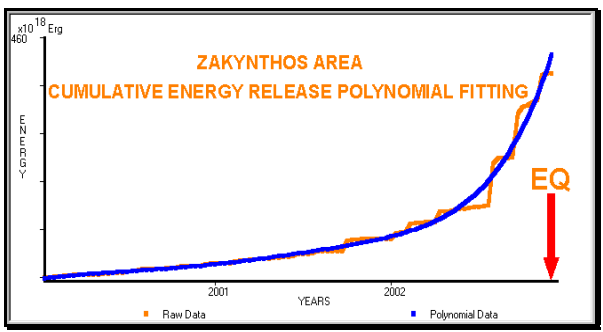

Fig. 3.1.4. The 6$^{th}$ order polynomial fitted, in the cumulative, seismic, energy release data for the area of Zakynthos. The brown line indicates the cumulative, energy data; the blue one indicates the fitted polynomial. The red arrow indicates the time when the Zakynthos EQ occurred.

The analytical expression of the polynomial which is obtained through the LSQ fitting procedure, utilizes the calculation of the time gradient of the polynomial and therefore, the rate of change of the cumulative, seismic energy release. This is presented in the following figure **(3.1.5)**.

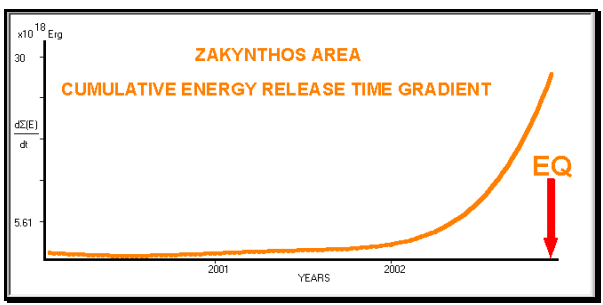

Fig. 3.1.5. Rate of change in time of the cumulative, seismic energy release observed, prior to the EQ of Zakynthos. The red arrow indicates the time when this EQ occurred.

It must be pointed out that all this analysis had been made before (one month) the EQ occurred and therefore, this EQ was expected to occur soon. The expected EQ did happen within a month and its location coincides with the location of the deep lithospheric fracturing which is mapped in this area **(fig. 3.1.6).**

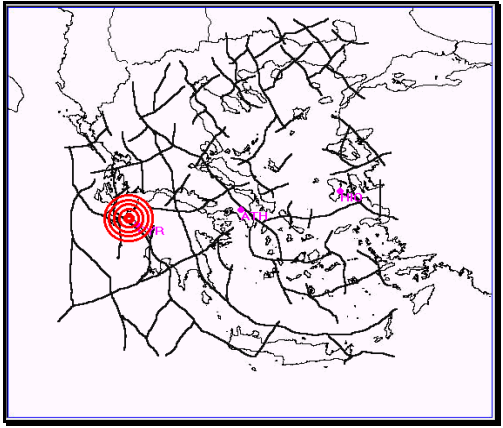

Fig. 3.1.6. Location of the EQ in Zakynthos in relation to the deep, lithospheric fracture zones, which are identified in the same area, by analyzing the gravity field.

thus verifying the validity of the corresponding, deep, lithospheric fracture zones / faults map (Thanassoulas, 1998).

### 3.2. Application of the theoretical model on the EQ in Kythira (08/01/2006, Ms = 6.9R).

This example is an "a posteriori" one. The validity of the method was tested against Kythira EQ (08/01/2006, Ms = 6.9R), which is presented in the following figure **(3.2.1).**

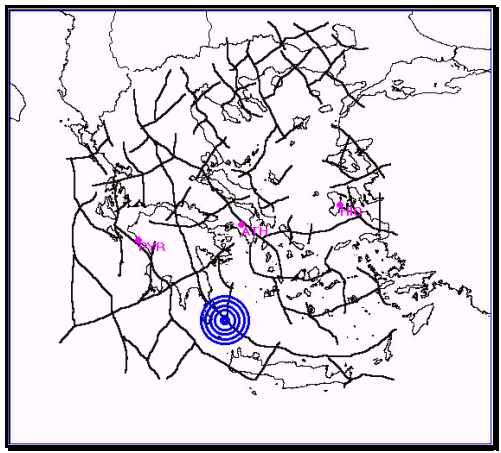

Fig. 3.2.1. Location (blue concentric circles) of the EQ in Kythira (08/01/2006, Ms = 6.9R), in relation with the location of the deep, fracture zones / faults of the lithosphere.

The seismogenic region, which is taken into account, is presented as a blue polygon frame in the figure **(3.2.2)** below.

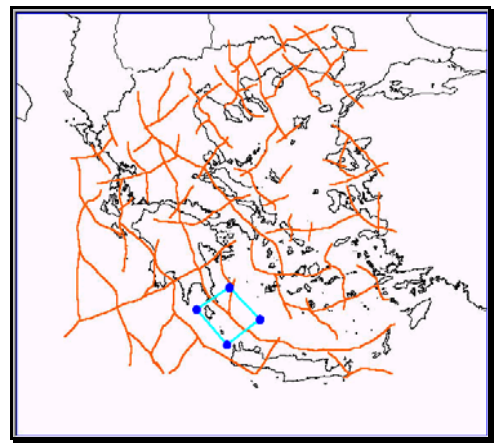

Fig. 3.2.2. Area of interest (blue frame) for which the lithospheric, seismic energy release methodology will be applied. Brown lines indicate the deep, lithospheric fracture zones.

The cumulative, seismic energy release, which is calculated for a period of 14 years (1992 – 2006), indicates that this seismogenic region was set in acceleration mode for the last 2 years (2004 – 2006). This is presented in figure **(3.2.3)**.

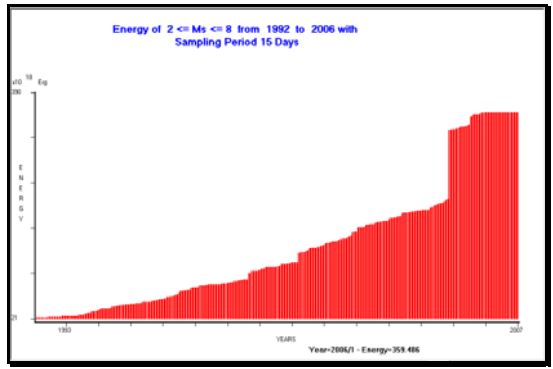

Fig. 3.2.3. Cumulative, seismic energy release determined, for the seismogenic region of Kythira, for the period 1992 – 2006.

The fitted, 6$^{th}$ order polynomial function of cumulative, released, seismic energy vs. time, indicates a rapid increase "graph knee", almost 2 years before the corresponding earthquake occurred. This is presented in next figure **(3.2.4)**.

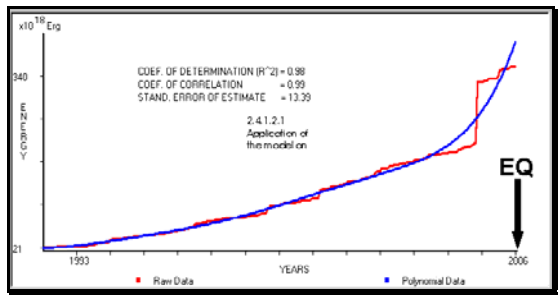

Fig. 3.2.4. Cumulative, seismic energy release is fitted by a 6$^{th}$ degree polynomial. The red line indicates the cumulative, seismic energy release data, while the blue one indicates the fitted, polynomial values. The black arrow indicates the time when the EQ occurred.

In this case, the time gradient, which is calculated from the analytical expression of the polynomial, resolves the time, when the seismogenic area entered the period of intense seismic acceleration in a much better way. The later is shown in next figure **(3.2.5)**.

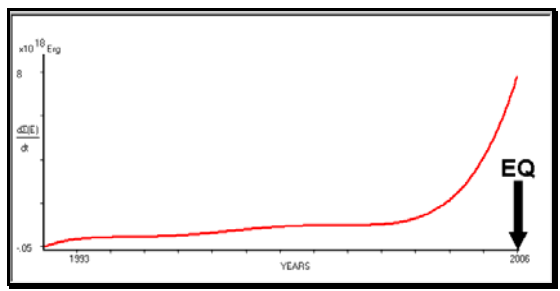

Fig. 3.2.5. Rate of change, in time, of the cumulative seismic energy release observed, prior to the EQ in Kythira. The black arrow indicates when the EQ occurred.

## 4. Discussion – conclusions.

A point that must be clarified, more, is the way the seismogenic area extent is set. What is accepted to date in the seismological community is that, the stronger the pending EQ is, the larger the area which is overcharged with strain deformation around the epicenter is. The later is represented by the following formula:

$$\log R = 0.42M - 0.68 \tag{4.1}$$

where: **(R)** is the radius of the strain charged area and **(M)** is the magnitude of the pending earthquake (Papazachos et al. 2000, Papazachos et al. 2001).

A radius of almost 200Km has been estimated, after having adopted this formula, for the assignment of the seismogenic region which is going to be used for the calculation of the seismic energy released for the case of the EQ in Kythira (**Ms = 6.9R**). As a first approach, this area is approximated by a surface, enclosed in a closely orthogonal frame of dimensions: 4 degrees (latitude) by 5 degrees (longitude). The later is presented in the following figure **(4.1)**.

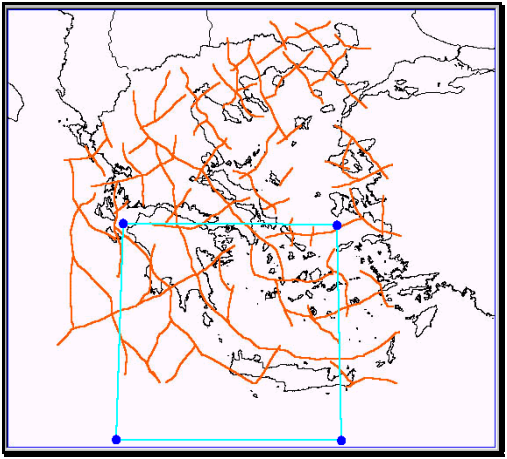

Fig. 4.1. Seismogenic area (blue frame) theoretically charged with strain deformation, required for the generation of the EQ in Kythira (Ms = 6.9 R).

It is more than evident that the seismogenic area, determined, by the use of equation **(4.1)** will be affected, not only by the main fault where this strong event took place, but also by the adjacent ones.

The later will contribute, with their seismic energy release, to the one, calculated for the entire frame. This must be avoided, when such calculations are made, since the seismic energy release of the fault which is expected to be activated, must only be studied. Moreover, calculations made for the identification of the accelerated deformation are masked by the presence of faults, which will not be activated at all. This is demonstrated in the following figures.

Let us assume a 2 by 2 degrees frame **(fig. 4.2)** that encloses some more main fracture zones, besides the one that was triggered.

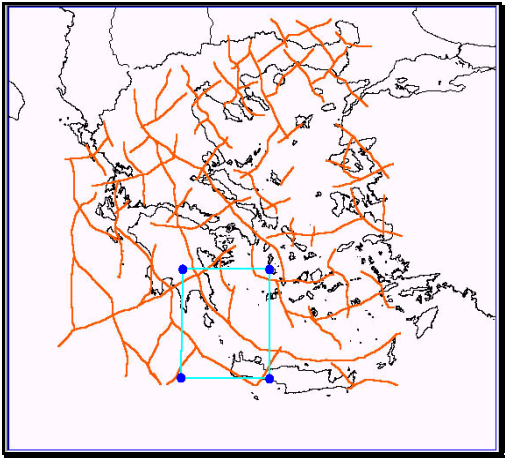

Fig. 4.2. A 2 by 2 degrees seismogenic area (blue frame), initially, assumed, for the cumulative, seismic energy release determination.

The cumulative seismic energy release is calculated as a function of time, for the same period (1992 – 2006) as the EQ of Kythira. This is presented in the next figure **(fig. 4.3).**

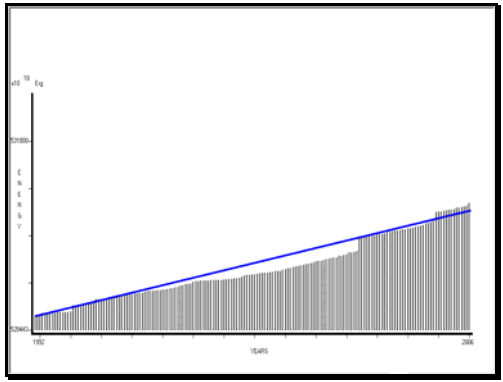

Fig. 4.3. The calculated cumulative, seismic energy release, for the same period (1992 – 2006) as for the EQ in Kythira.

This graph fits quite well a straight line, except for the period 1998-2002, when a moderate magnitude seismic event took place.

There is no evidence to show that any, accelerated, deformation mechanism has been initiated. The same is obvious from the next figure **(fig. 4.4)**.

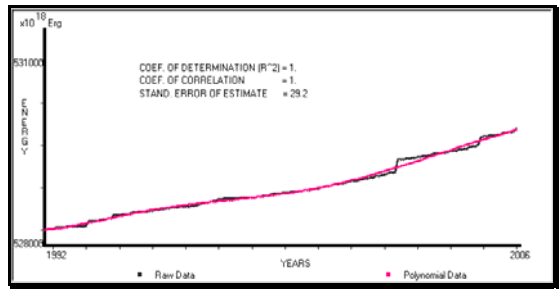

Fig. 4.4. Polynomial fitting (6<sup>th</sup> order), as determined on the cumulative seismic energy release data.

For the next example, a smaller frame (1 by 1 degree) was used, which encloses only the activated fault area, but does not take into account the strike of the seismically, active fault. This frame is presented in next figure **(4.5)**.

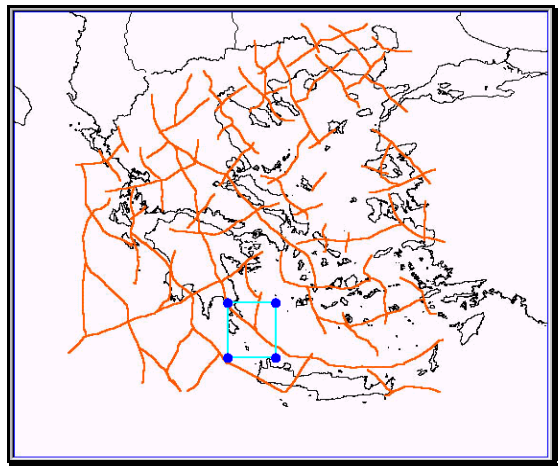

Fig. 4.5. A 1 by 1 degrees seismogenic area (blue frame), initially, assumed, for the cumulative, seismic energy release determination.

The corresponding, cumulative, seismic energy release graph is shown in the following figure **(4.6)**.

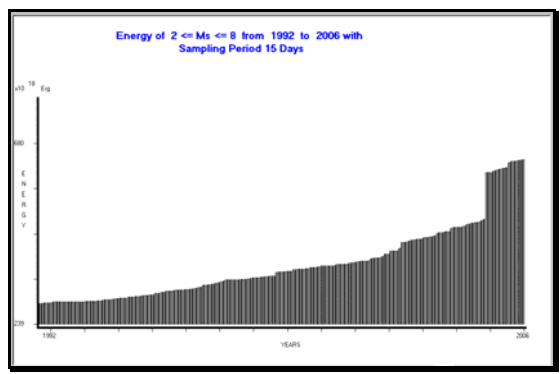

Fig. 4.6. Cumulative, seismic energy release, determined for the 1 by 1 degrees seismogenic region, for the period 1992 – 2006.

The accelerating deformation characteristics of the cumulative seismic energy release are initially revealed in this graph. This is more obvious in the following figure **(4.7)** where the polynomial, fitting, procedure was used.

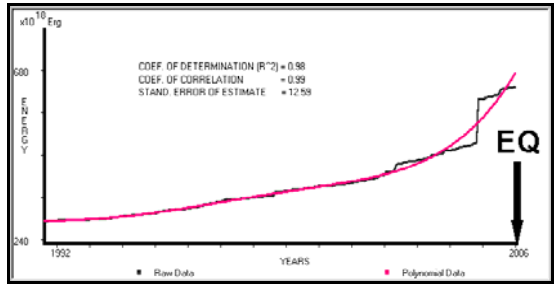

Fig. 4.7. Polynomial fitting (6$^{th}$ order) as determined from the cumulative seismic energy release data.

The same accelerating deformation character of the seismicity of the corresponding frame is indicated in more details by the time gradient graph **(fig. 4.8)** of the cumulative seismic energy release data.

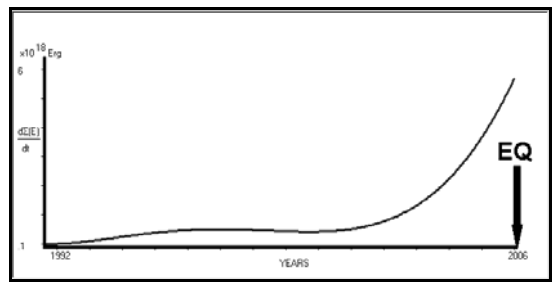

Fig. 4.8. Rate of change in time observed, for the corresponding 1 by 1 degrees seismogenic area, of the cumulative, seismic, energy release.

The comparison of the three cases presented (2 by 2, 1 by 1, and frame oriented along fault strike) indicates that, the most indicative case, for the detection of the accelerated deformation status of a seismogenic region, is the one, where the assumed regional seismogenic area (frame) conforms to the tectonic character of the seismogenic area and the strike of the seismogenic fault.

The later leads us to the following conclusion.

In the wider regional area, where a strong EQ will take place, the equation:

$$logR = 0.42M - 0.68 \qquad (4.2)$$

is valid and generally defines the strain-deformed, area.

However, in the vicinity of the fault, to be activated, the most drastic, elastic deformations take place and this explains why the EQ takes place in this fault, during the strain release. Consequently, this leads directly to the "Rebound Theory of Reid".

After having presented the above examples, what is proposed and is more probable for the seismogenic area is the following:

- The LSEFM considered as an open physical system that represents any seismogenic region justifies through its mathematical analysis the observations made by the seismologists and referred to as "accelerated deformation" and "seismic quiescence".

- Up to a distance **(R)** from the epicenter location, calculated by the equation (**4.2**), during the preparation of a strong EQ, strain is accumulated.

- Very close to the fault which will be activated, the elastic deformation prevails and the dominant model for the earthquake generation mechanism is the one of the Rebound Theory.

It is obvious that, the accelerated deformation status of each deep, lithospheric fracture zone must be studied in detail. Further more, it is possible to compile maps which will indicate the acceleration deformation spatial distribution all over a wide seismogenic area, and hence the corresponding, seismic potential / risk can be evaluated.

In conclusion, the combined use of the lithospheric fracture zones and the cumulative, seismic energy flow lithospheric model selected for detailed analysis can provide us with valuable information for the seismic energy charge of any area.

Finally, the presented mathematical analysis (**equation 2.8**) sets the basis for the determination of the maximum expected magnitude of an earthquake that will occur in the future in an already seismically activated area. Moreover, it is possible to compile "seismic potential maps" of regional seismological interest and seismic hazard mitigation. An extended presentation of this topic was given by Thanassoulas (2007).

In the next two articles to come soon, the magnitude determination and the seismic potential will be presented.

## 5. References.